\documentclass[a4paper,11pt]{article}

\usepackage{kpfonts}
\usepackage[utf8]{inputenc}
\usepackage[margin=3cm]{geometry}

\usepackage{graphicx}
\usepackage{subcaption}
\usepackage{amsmath,amssymb,amsthm}
\usepackage{mathtools}
\usepackage{thm-restate}
\usepackage{microtype}
\usepackage{enumitem}
\usepackage{xcolor}
\usepackage{xspace}
\usepackage[nocompress]{cite}
\definecolor{linkblue}{HTML}{2B4C7E}
\usepackage[pdfpagelabels,colorlinks,allcolors=linkblue]{hyperref}
\usepackage[nameinlink,capitalize,noabbrev,sort]{cleveref}
\hypersetup{
    pdftitle={Tripod Reordering: A 33-Queue Layout of Planar Graphs},
    pdfauthor={Sergey Pupyrev}
}

\DeclareMathOperator{\qn}{qn}

\theoremstyle{plain}

\newtheorem{lemma}{Lemma}
\newtheorem{claim}{Claim}

\graphicspath{{tex/pics/}{pics/}}

\newcommand{\Qm}{Q_{\mathrm m}}
\newcommand{\Qt}{Q_{\mathrm t}}
\newcommand{\Qb}{Q_{\mathrm b}}
\newcommand{\HH}{{\ensuremath{\widehat H}}}
\newcommand{\HHtri}{{\ensuremath{\HH^{\triangle}}}}
\newcommand{\KK}{{\ensuremath{\widehat K}}}

\definecolor{brightmaroon}{rgb}{0.76, 0.13, 0.28}

\newcommand{\df}[1]{\textcolor{brightmaroon}{\emph{#1}}}

\newcommand{\alg}[1]{\textsc{#1}}
\newcommand{\pps}{\alg{PPS}\xspace}
\newcommand{\ptt}{\alg{P3T}\xspace}

\begin{document}

\title{Tripod Reordering: A 33-Queue Layout of Planar Graphs}
\author{
    Sergey Pupyrev \\
    spupyrev@gmail.com
}

\date{}
\maketitle

\begin{abstract}
    A \emph{queue layout} of a graph consists of a linear order of the vertices and a partition
    of the edges into queues so that no two edges in a single queue are nested. The minimum number of queues
    needed in a queue layout of a graph is called its \emph{queue number}. A
    planar product structure theorem of
    Dujmovi{\'c}, Joret, Micek, Morin, Ueckerdt, and Wood (JACM, 2020)
    states that every planar graph
    is a subgraph of the strong product of a graph of simple treewidth at most $3$, a clique $K_3$, and a path.
    The result has been used to show that the queue number of planar graphs is at most $49$.

    Recently, Bekos, Gronemann, and Raftopoulou (Algorithmica, 2023) investigated how the general approach based on the product structure can be
    optimized for planar graphs. They claim that an appropriate reordering of the three vertices
    associated with each bag in the product structure decomposition,
    called a \emph{tripod}, reduces the queue number of planar graphs to $42$.
    We first highlight a gap in their queue layout algorithm: one of the
    choices required by the algorithm is not guaranteed to exist.
    Then we show how to address the limitation and provide an alternative
    tripod reordering mechanism that reduces the upper bound on the
    queue number of planar graphs to $33$.
    As a corollary, bipartite planar graphs have queue number
    at most $22$.
\end{abstract}

\section{Introduction}
Let $G=(V,E)$ be a simple undirected graph, and let $\sigma$
be a linear order of its vertices. We write $u\prec_\sigma v$ if $u$ precedes $v$ in $\sigma$, and $u\preceq_\sigma v$ if $u\prec_\sigma v$ or $u=v$.
Two edges $(u, v) \in E$ and $(x, y) \in E$ are \df{nested} with respect to $\sigma$
if $u\prec_\sigma x\prec_\sigma y\prec_\sigma v$ or $x\prec_\sigma u\prec_\sigma v\prec_\sigma y$. A set of edges of $G$ forms a \df{queue} with respect to $\sigma$
if no two edges in the set are nested in $\sigma$. Given a linear order $\sigma$ of the vertices of $G$, the edges
$(u_1, v_1), \dots, (u_k, v_k)$ form a \df{rainbow} of size $k$ if $u_1\prec_\sigma\cdots\prec_\sigma u_k\prec_\sigma v_k\prec_\sigma\cdots\prec_\sigma v_1$.
The edges of $G$ can be partitioned into $k$ queues if and only if there is no rainbow of size $k+1$ in $\sigma$~\cite{HR92}.
A \df{queue layout} of $G$ consists of a linear order
of its vertices and a partition of its edges into queues
with respect to that order. The minimum number of queues needed in a queue layout of $G$
is called its \df{queue number}, denoted by $\qn(G)$.

Queue layouts were introduced by Heath and Rosenberg~\cite{HR92} and have been studied extensively over the
years~\cite{HLR92,BFP13,Pem92,DF15,DW05,DMW05,Wie17,Pup17,ABGKP20,FKMPR23,Katheder0PU25,AlamBGKP22,Pup22,ABGKP23}.
Several basic graph classes have bounded queue number.
Every tree has queue number one~\cite{HR92}, outerplanar graphs have queue number at most two~\cite{HLR92}, and series-parallel graphs have queue number at most three~\cite{Wie17}. Alam, Bekos, Gronemann, Kaufmann, and Pupyrev~\cite{ABGKP20} showed that the queue number of
planar $3$-trees is at most five; they also gave a lower bound of four.
The central problem in the field for many years was whether planar graphs have bounded queue number.
Heath, Leighton, and Rosenberg~\cite{HLR92} conjectured that this is the case.
Di~Battista, Frati, and Pach~\cite{BFP13} and Dujmovi{\'c}~\cite{Duj15} obtained logarithmic upper bounds.
Subsequently, Bekos, F{\"{o}}rster, Gronemann, Mchedlidze, Montecchiani, Raftopoulou, and
Ueckerdt showed that planar graphs of bounded maximum degree have constant
queue number~\cite{BFGMMRU19}.
Finally in 2020, Dujmovi{\'c}, Joret, Micek, Morin, Ueckerdt, and Wood
settled the conjecture by showing that the queue number of every planar graph is bounded~\cite{DJMMUW20}.
Their proof uses the planar product structure theorem: every planar graph is a subgraph of the strong product of a graph of simple treewidth at most $3$, a clique $K_3$, and a path.
Combining this theorem with Wood's bounds for queue layouts of graph products~\cite{W05} gives the constant $49$.

Bekos, Gronemann, and Raftopoulou~\cite{BGR23} revisited the approach based on the product structure and claimed an improvement from $49$ to $42$ queues for
planar graphs.
Their argument uses the freedom to reorder the three vertices associated with each \emph{tripod} (a bag in the product structure decomposition);
we refer to this optimization as \df{tripod reordering}.
Unfortunately, it turned out that one of the choices required by their ordering rule is not guaranteed to exist. Consequently, the published proof in~\cite{BGR23} does not establish the claimed upper bound of $42$ queues for planar graphs.\footnote{Although the proof in~\cite{BGR23} does not establish
the $42$-queue bound, its reduction of the intra-bag
contribution by one queue yields an upper bound of $48$.
This observation also appears in~\cite{DJMMUW20}
and is used in our construction.}
We explain the gap in \cref{sec:gap}, and also refer to our note with a concrete
counter-example~\cite{PupGap26}.

In this paper, we revisit the tripod reordering technique that can be used for
building queue layouts of planar graphs. As in~\cite{BGR23}, the approach is based on the planar product decomposition algorithm, which we discuss in
detail in \cref{sec:tripod-decomposition}.
The main difference is that we work directly with a multigraph
arising from the recursive decomposition, avoiding the problematic
ordering step. The key observation is that the resulting vertex
order is compatible with the rotation system of this multigraph.
This compatibility, defined formally in \cref{sec:p3t-properties},
allows us to save more queues by reordering the tripods,
yielding the following improved upper bound.

\begin{restatable}{theorem}{maintheorem}
    \label{thm:main}
    Every planar graph has queue number at most \(33\).
\end{restatable}

We provide a full implementation of the $33$-queue layout algorithm in~\cite{bob}.

Following the work of F{\"o}rster, Kaufmann, Merker, Pupyrev,
and Raftopoulou~\cite{FKMPR23}, we can transfer our upper bound
to bipartite planar graphs. The same algorithm applies,
with only a change in the counting of queues.

\begin{restatable}{theorem}{bitheorem}
    \label{thm:bipartite}
    Every bipartite planar graph has queue number at most \(22\).
\end{restatable}

The rest of the paper is organized as follows.
\Cref{sec:framework} reviews the planar product decomposition
and the queue layout algorithm for planar $3$-trees.
We then explain the gap in the previous tripod reordering
in \cref{sec:gap} and outline our construction in
\cref{sec:overview}. \Cref{sec:bottom-parent-33} presents
the construction in detail and proves \cref{thm:main}.
Finally, in \cref{sec:conclusions}, we derive the bipartite bound
of \cref{thm:bipartite} and discuss open problems.

\section{Preliminaries}
\label{sec:framework}

In this section, we introduce basic definitions and tools that we use to investigate  the queue number of planar graphs.

\subsection{The Product Structure Approach}
\label{sec:layered-partitions}

The \df{strong product} $ G_1 \boxtimes G_2 $ of two graphs $ G_1 $ and $ G_2 $ is a graph with vertex set $ V(G_1) \times V(G_2) $ and an edge between two vertices $ (v_1, v_2) $ and $ (w_1, w_2) $ if
\begin{enumerate}
    \item  $ v_1 = w_1 $ and $(v_2,w_2) \in E(G_2) $,
    \item $ v_2 = w_2 $ and $(v_1,w_1) \in E(G_1) $, or
    \item $(v_1,w_1) \in E(G_1) $ and $(v_2,w_2) \in E(G_2) $.
\end{enumerate}

Let $G$ and $H$ be graphs. An \df{$H$-partition} of $G$ is a partition
$\{A_x : x\in V(H)\}$ of $V(G)$ into bags such that, for every edge
$(u,v)\in E(G)$ with $u\in A_x$ and $v\in A_y$, either $x=y$ or
$(x,y)\in E(H)$. In the former case the edge is \df{intra-bag}; in the latter
case it is \df{inter-bag}. A \df{layering} of $G$ is a partition
$(V_0,V_1,\ldots)$ of $V(G)$ according to the shortest distances from a fixed root.
An edge is \df{intra-layer} if its endpoints lie in the same layer, and
\df{inter-layer} otherwise. The $H$-partition has \df{layered width} $\ell$
with respect to the layering if $|A_x\cap V_i|\le \ell$ for every
$x\in V(H)$ and every layer $V_i$.

We use the strong product representation of an $H$-partition
to bound the queue number of $G$.
If the partition has
layered width $\ell$ with respect to a layering,
then $G$ is a subgraph of $H\boxtimes K_\ell\boxtimes P$
for a path $P$~\cite{DJMMUW20}. The path records the layers, while
the clique records the positions of vertices within
each intersection of a bag and a layer.
The queue layout of $G$ is constructed in two steps.
Start with a queue layout of $H$.
The first step is to replace each vertex of $H$ by $\ell$ consecutive
vertices to obtain an order of $H\boxtimes K_\ell$.
If the original layout of $H$ admits $\qn(H)$ queues, edges between
distinct replacement cliques use at most $\ell \cdot \qn(H)$ queues,
and edges within the cliques use at most
$\lfloor\ell/2\rfloor$ additional queues.
Thus $\qn(H\boxtimes K_\ell)\le
\ell\cdot\qn(H)+\lfloor\ell/2\rfloor$.
The second step is to take the strong product of the result with a path $P$.
The resulting vertex order lists the layers (that is, copies of $H\boxtimes K_\ell$)
consecutively. Within each layer, it follows the order
of $H\boxtimes K_\ell$: bags occur in the vertex order
of $H$, and the vertices of each bag occur consecutively.
For a product with a path, the construction uses three
queues for each queue of the other factor, together with
one queue containing all edges between copies of the
same vertex. Restricting the resulting layout to $G$
gives a queue layout of $G$.
The two product bounds are summarized below.

\begin{claim}[Theorem~7 of \cite{W05}, Lemma~9 of \cite{DJMMUW20}]
    \label{lem:product-queues}
    For every graph $H$, path $P$, and integer $\ell\geq1$,
    \[
        \qn(H\boxtimes P)\le 3\qn(H)+1,
        \qquad
        \qn(H\boxtimes K_\ell)
        \le \ell\cdot\qn(H)+\left\lfloor\frac{\ell}{2}\right\rfloor.
    \]
\end{claim}

\begin{figure}[!tb]
    \centering
    \begin{subfigure}[c]{.48\linewidth}
        \centering
        \includegraphics[page=1,width=\linewidth]{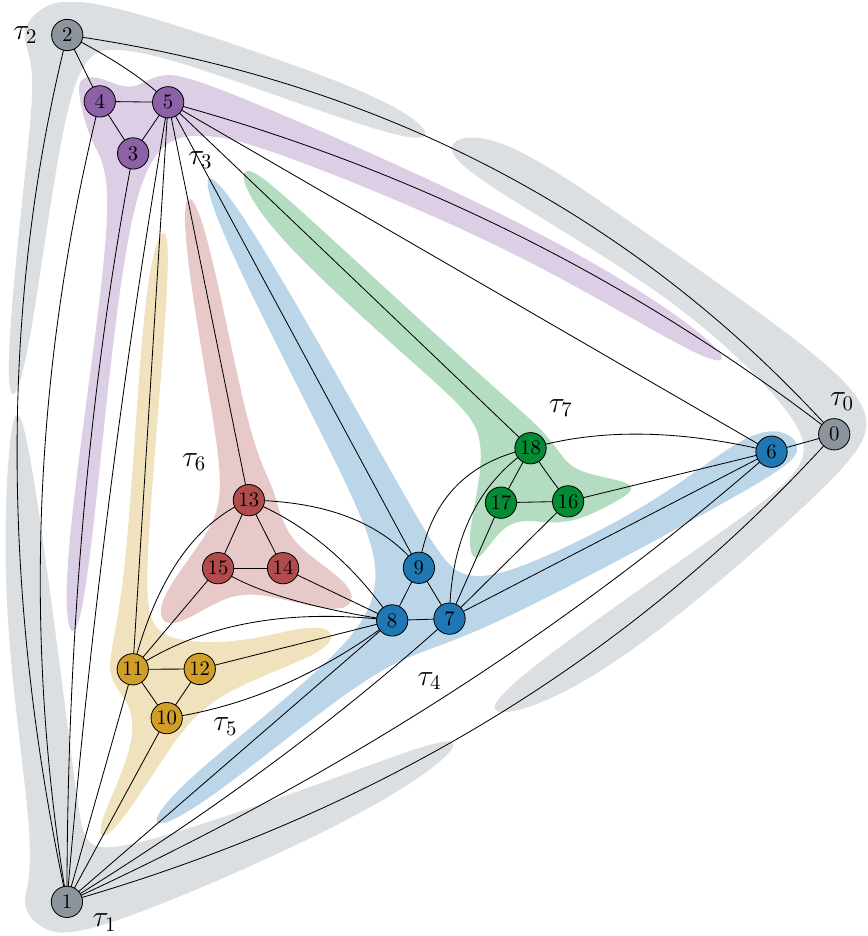}
        \caption{The graph and its tripod partition}
    \end{subfigure}
    \hfill
    \begin{subfigure}[c]{.48\linewidth}
        \centering
        \includegraphics[page=2,width=\linewidth]{decomposition_overview}
        \caption{The recursive regions around $\tau_4$}
    \end{subfigure}
    \caption{Tripod decomposition. (a) The graph and its tripod partition.
        (b) The regions $R^1,R^2,R^3$ around $\tau_4$, with paths
        $p^1=(6,7)$, $p^2=(8)$, $p^3=(9)$.
        Tripods in $R^j$ have no neighbors on $p^j$;
        $\tau_6$ lies in $R^1$ and $\tau_7$ in $R^2$.}
    \label{fig:decomposition-review}
\end{figure}

To apply \cref{lem:product-queues} to planar graphs, the algorithm
of~\cite{DJMMUW20} constructs, for a given planar graph $G$, an $H$-partition of layered width $3$
with respect to a layering, where $H$ is a partial planar $3$-tree.
This is the planar product structure theorem; we denote the
corresponding procedure by the \df{\pps algorithm}.
In order to build a queue layout for a planar graph, \pps needs a low-queue
layout of $H$. This is done by applying (in a black-box manner) the algorithm
of~\cite{ABGKP20}, referred to as the \df{\ptt algorithm},
which provides a $5$-queue layout for planar $3$-trees. Combining the results,
we apply the clique bound in \cref{lem:product-queues} with $\ell=3$
and $\qn(H)\le5$ to obtain
$\qn(H\boxtimes K_3)\le3\cdot5+\lfloor3/2\rfloor=16$.
Applying the path bound to $H\boxtimes K_3$ then gives
$\qn(G)\le\qn\big((H\boxtimes K_3)\boxtimes P\big) \le 3\cdot16+1=49$.

Next we provide details for the two algorithms, \pps and \ptt, since both are essential for the proof of \cref{thm:main}.

\subsection{Tripod Decomposition of Planar Graphs}
\label{sec:tripod-decomposition}

The \pps algorithm builds an $H$-partition of a planar graph $G$
recursively. Assume that $G$ is a plane triangulation, and fix a
BFS tree rooted at a vertex of the outer face.
A \df{vertical path} is a subpath of a root-to-vertex path in
this tree; it meets each layer at most once.
A \df{tripod} consists of at most three pairwise disjoint
vertical paths whose endpoints farthest from the root
are pairwise adjacent in $G$.
Each of the three vertices of the outer face forms an
\df{initial} tripod consisting of one vertex.
\pps recursively decomposes regions bounded by at most
three tripods.
In a region bounded by tripods $\tau_1,\tau_2,\tau_3$, the algorithm
partitions the boundary into three nonempty consecutive paths,
each contained in one boundary tripod, and assigns them colors
$1,2,3$. When fewer than three boundary tripods occur, their
boundary paths are split to obtain three color classes.
Each interior vertex receives the color of the first boundary
vertex on its tree path toward the root.
Sperner's Lemma~\cite{AZ10} guarantees a triangular face whose vertices
have three distinct colors.

For each vertex of this face, the algorithm follows the
path in the BFS tree toward the root until reaching the
first boundary vertex, then removes that boundary endpoint. The remaining paths, denoted by $p^1,p^2,p^3$,
together with the edges of the selected face,
form the new tripod $\tau$.
The face and the paths divide the region into at most three
smaller regions, each bounded by at most three tripods from
$\tau,\tau_1,\tau_2,\tau_3$.
These regions are denoted by $R^1,R^2,R^3$; see
\cref{fig:decomposition-review}(b). Tripods created inside
$R^j$ have no edges to $p^j$.
The algorithm recurses on the remaining nonempty regions.
The resulting tripods partition $V(G)$; see
\cref{fig:decomposition-review}(a).
Graph $H$ has one vertex for each tripod, and two
vertices of $H$ are adjacent whenever an edge of $G$ joins
their tripods. We call $H$ the \df{quotient} of the tripod
partition. Since each tripod contains at most three
vertical paths, this gives an $H$-partition of layered width
at most $3$. The construction guarantees that $H$ is a partial
planar $3$-tree.

To build the queue layout of $G$, \pps augments $H$ to a
maximal plane $3$-tree and applies the \ptt algorithm to obtain
a $5$-queue layout. The two product bounds in
\cref{lem:product-queues} then give a $49$-queue layout
of $G$. The next subsection describes \ptt.

\subsection{The P3T Queue Layout Algorithm}
\label{sec:p3t}

The original \ptt algorithm, introduced in \cite{ABGKP20},
constructs a $5$-queue layout of a (partial) planar $3$-tree,
denoted by $H$. As a preprocessing step, $H$ is augmented to a
maximal plane $3$-tree. Its vertices are then partitioned into
\df{levels} according to their shortest distance from the outer face;
we use \emph{levels} for $H$ to distinguish them from the \emph{layers} of $G$.
The first level consists of the three vertices incident to the
outer face; each subsequent level consists of the vertices
incident to the outer face after deleting all preceding levels.
A \df{level component} is a connected component of the subgraph
induced by one level of $H$.
If $H$ is a maximal plane $3$-tree, then every level component is internally triangulated
and outerplane~\cite{ABGKP20}.
Each component beyond the first
level lies inside a triangular face of a component in the
preceding level, called its \df{parent component}; its neighbors
in that level are precisely the three vertices of the containing
face.
These parent-child relations form a tree whose root is
the component of the first level.
The \ptt algorithm orders the components by a breadth-first
traversal of this tree, ordering the children of each component
by their containing faces. Replacing each component in this
traversal by its vertex order gives the vertex order of $H$.
Vertices appear level by level, with each level component
consecutive, and the construction reduces to ordering an
outerplane component and its children.

\begin{figure}[tb]
    \centering
    \includegraphics[page=3,width=.8\linewidth]{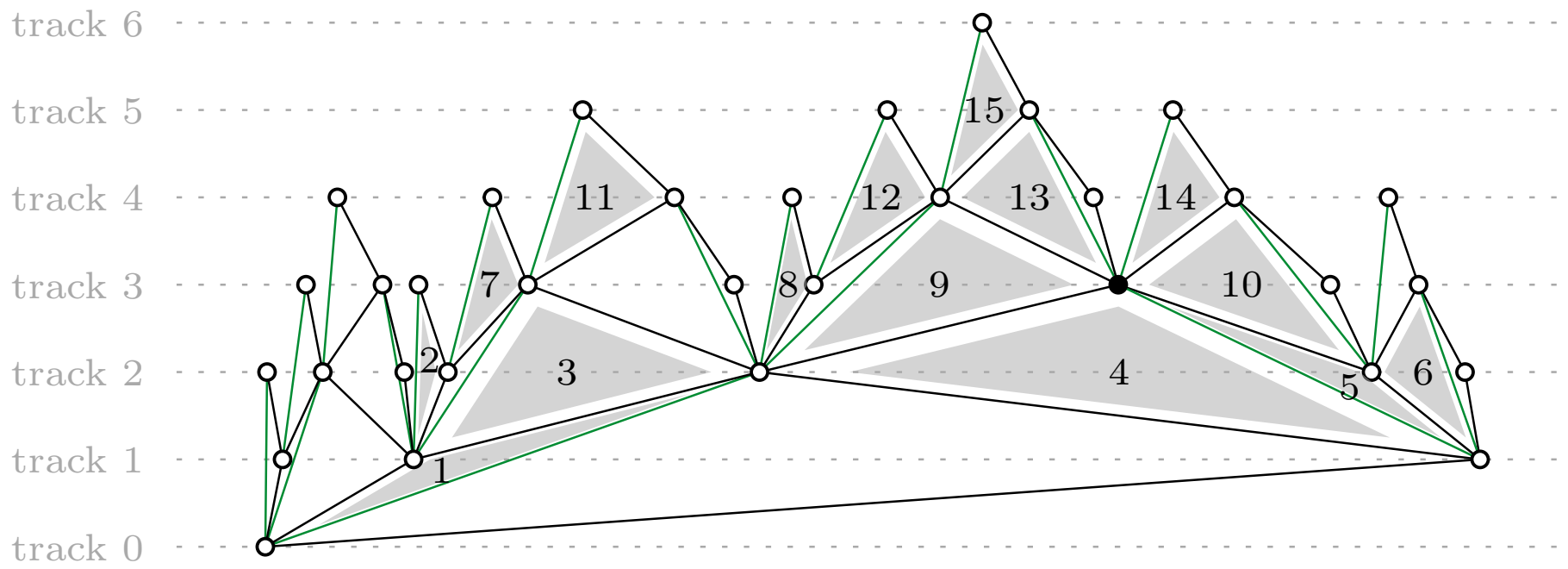}
    \caption{The two-level construction in the \ptt algorithm.
        Black and green edges have spans one and two and belong to
        $Q_1$ and $Q_2$, respectively. Shaded triangles represent child
        components, numbered $1,\ldots,15$ by comparing their containing faces:
        first by the bottom vertex, then by the middle vertex,
        and finally by the top vertex.}
    \label{fig:out-overview}
\end{figure}

This two-level construction is illustrated in \cref{fig:out-overview}.
It relies on a $2$-queue layout of an outerplane graph $K$, described by
Dujmovi{\'c}, P{\'o}r, and Wood~\cite{DPW04} in the context of
\emph{track layouts}. The layout is best described via a straight-line
outerplane drawing of $K$ in which vertices are placed on
horizontal \df{tracks} at integer heights. The \df{span} of an edge
is the absolute difference between the heights of its endpoints.
In this drawing, every edge has span one or two. The algorithm
orders the vertices by increasing height, breaking ties by increasing
horizontal coordinate. The span-one edges and the span-two edges
each form a queue. Starting from an outer-face edge whose endpoints
are placed on consecutive tracks, we add each new triangle above
the current drawing, using the construction of~\cite{DPW04}.
Thus the upward vertical ray from every vertex
lies in the outer face.
The vertices of each
triangular face have consecutive heights and are called its
\df{top}, \df{middle}, and \df{bottom} vertices. The algorithm
orders the child components lexicographically
by the triples (bottom, middle, top) of vertices of their
containing faces, using the parent vertex order for each comparison.
The same drawing construction
determines the vertex order within every child component.
\Cref{thm:outerplane-drawing} describes the properties of the
obtained drawing.

\begin{claim}[Lemma 22 of \cite{DPW04}]
    \label{thm:outerplane-drawing}
    Every connected outerplane graph has a straight-line
outerplane drawing respecting its embedding in which every
edge has span one or two and the upward vertical ray from
every vertex lies in the outer face. Ordering the vertices
track by track from bottom to top, and from left to right
within each track, makes the span-one edges and the span-two
edges each form a queue.
\end{claim}

The \ptt algorithm applies the drawing of
\cref{thm:outerplane-drawing} to every level component of $H$.
Edges joining vertices in the same level are called
\df{level} edges; those of span one and two use $Q_1$
and $Q_2$, respectively. Edges joining consecutive levels
are called \df{binding} edges; they use three queues,
denoted $\Qt$, $\Qm$, and $\Qb$.
At a parent vertex $u$, binding edges to a child component use
$\Qt$ if its containing face lies above $u$, $\Qb$ if it lies
below $u$, and $\Qm$ if $u$ is its middle vertex.
Within each binding queue, planarity ensures that the
parent endpoints occur in the same order as the containing
faces, and hence as the child components. Thus two edges
with distinct parent endpoints cannot nest; edges with
the same parent endpoint do not nest either.
Since concatenating level components preserves the level queues
and the breadth-first order of the level components preserves the binding queues,
the five sets $Q_1,Q_2,\Qt,\Qm,\Qb$ form a $5$-queue layout of $H$.

\section{Tripod Reordering: Old and New}

\subsection{A Gap in the Previous Construction~\cite{BGR23}}
\label{sec:gap}

Bekos, Gronemann, and Raftopoulou~\cite{BGR23} refine the
product structure approach described in \cref{sec:layered-partitions}
by choosing the order of vertices within each bag.
Every bag has at most three vertices in a layer, one
from each vertical path of its tripod. Instead of ordering
these vertices arbitrarily, their algorithm chooses an order
of the three vertical paths and uses it consistently in all
layers. The inter-bag edges are divided into three types:
intra-layer edges and inter-layer edges in the two directions
of the vertex order of $H$. Without reordering, each type
uses $\ell\cdot\qn(H)\le3\cdot5=15$ queues, since the layered
width is $\ell=3$ and $H$ has a $5$-queue layout.
The tripod reordering is intended to save two queues
for each type, reducing the count from $15$ to $13$.
Together with a separate improvement for intra-bag edges,
this gives the claimed bound of $3\cdot13 + 3 = 42$ queues.

The reordering uses the observation that tripods created in
the same recursive region have no edges to one common vertical
path of the boundary tripod. This follows from the construction
in \cref{sec:tripod-decomposition} and is also stated in
\cref{lem:new-structure}(iii).
Consider the tripod $\tau_4$ in \cref{fig:decomposition-review}(b).
Its three vertical paths $p^1,p^2,p^3$ divide the region
in which it was created into three recursive regions
$R^1,R^2,R^3$. Tripods subsequently created inside $R^j$
have no edges to $p^j$. Thus all tripods in one recursive region
miss the same path of $\tau_4$, where a set of vertices
\df{misses} a path if no edge joins the set to that path.
The ordering rule in~\cite{BGR23} uses this property to
choose the first vertical path of a tripod $\tau$: for certain
level components in the next level of $H$,
it requires a path missed by all vertices of $G$ belonging
to the tripods represented in that component. The remaining
two paths are ordered arbitrarily.
However, the existence of a common missed path is guaranteed
only for tripods in the same recursive region.
Before applying \ptt, the algorithm subdivides loops
and parallel edges in the quotient and augments the resulting
simple graph to a maximal plane $3$-tree. This augmentation
can join tripods from different recursive regions into one
next-level component, violating the condition required by
the ordering rule.

\begin{figure}[!tb]
    \centering
    \begin{minipage}[c]{.49\linewidth}
        \begin{subfigure}{\linewidth}
            \centering
            \includegraphics[page=1,width=\linewidth]{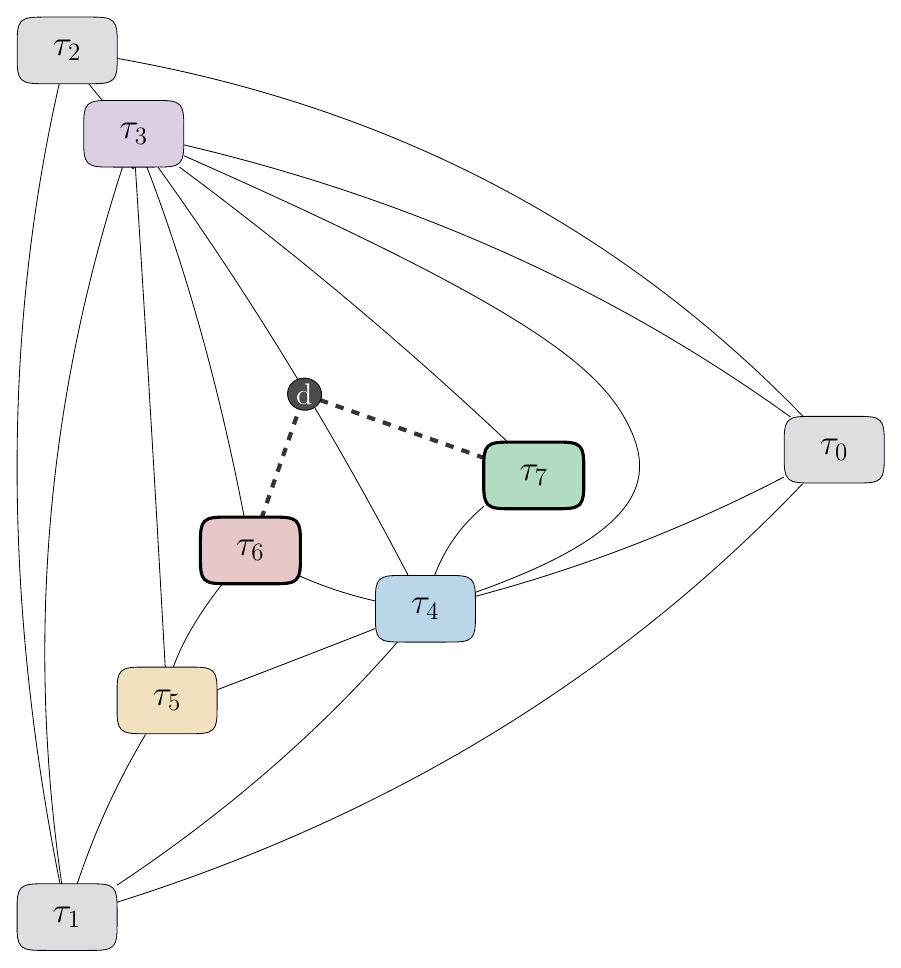}
            \caption{The augmented quotient}
        \end{subfigure}
        \par\medskip
        \begin{subfigure}{\linewidth}
            \centering
            \includegraphics[page=2,width=\linewidth]{gap_overview}
            \caption{The adjacencies to the paths of $\tau_4$}
        \end{subfigure}
    \end{minipage}
    \hfill
    \begin{minipage}[c]{.49\linewidth}
        \begin{subfigure}{\linewidth}
            \centering
            \includegraphics[page=3,width=\linewidth]{gap_overview}
            \caption{The multigraph quotient $\HH$}
        \end{subfigure}
        \par\medskip
        \begin{subfigure}{\linewidth}
            \centering
            \includegraphics[page=4,width=\linewidth]{gap_overview}
            \caption{Tripods arranged by quotient levels}
        \end{subfigure}
    \end{minipage}
    \caption{The gap in the previous tripod reordering~\cite{BGR23}
        for the graph in \cref{fig:decomposition-review}.
        (a)~Augmentation edges through $d$ place $\tau_6,\tau_7$ in one next-level component.
        (b)~Together, $\tau_6,\tau_7$ have neighbors on all three paths of $\tau_4$.
        (c--d) Retaining loops and parallel edges keeps $\tau_6,\tau_7$ in separate components.
        In (d), intra-tripod edges are omitted.}
    \label{fig:gap-review}
\end{figure}

\Cref{fig:gap-review}(a--b) illustrates this problem for the
graph in \cref{fig:decomposition-review}. The quotient has
three levels, $L_0=\{\tau_0,\tau_1,\tau_2\}$,
$L_1=\{\tau_3,\tau_4,\tau_5\}$, and $L_2=\{\tau_6,\tau_7\}$;
see \cref{fig:gap-review}(d).
The augmentation shown in \cref{fig:gap-review}(a) adds
a vertex $d$ to $L_2$ and edges $(\tau_6,d)$ and $(d,\tau_7)$,
joining $\tau_6$ and $\tau_7$ into one level component.
However, these tripods lie in different recursive regions
of $\tau_4$: $\tau_6$ misses $p^1$ and $\tau_7$ misses $p^2$.
Together they have neighbors on all three paths of $\tau_4$,
as shown in \cref{fig:gap-review}(b), so the ordering rule
cannot choose the required first path. Our note~\cite{PupGap26} presents
the full $19$-vertex example
and explains the gap in the algorithm.

\subsection{New Construction}
\label{sec:overview}

Our approach follows the same product structure decomposition, using the \pps algorithm.
To avoid the limitation of~\cite{BGR23}, we retain the loops
and parallel edges created by contracting the tripods.
The resulting embedded multigraph is denoted by $\HH$
(throughout the paper, the notation~$\widehat{\cdot}$
distinguishes multigraphs from simple graphs).
These parallel edges separate the recursive regions; for example,
in \cref{fig:gap-review}(c), the parallel edges keep $\tau_6$
and $\tau_7$ in separate next-level components.
We extend the \ptt algorithm to work directly with $\HH$, avoiding
the subdivisions and augmentation used in the previous
construction; \cref{fig:gap-review}(d) illustrates the
resulting order of the level components.

Extending \ptt to $\HH$ gives a $5$-queue layout, so
$\qn(\HH)\le5$. The key observation is that the vertex order
in this layout is compatible with the rotation system of
$\HH$: within each queue, the order of neighboring vertices
or child components agrees with their cyclic arrangement
around a vertex. This compatibility, defined formally in
\cref{sec:p3t-properties}, allows us to choose the order
of the three vertical paths of each tripod so as to reduce
the number of queues needed after expansion.
Replacing each vertex of $\HH$ by one vertex for each
vertical path of its tripod gives an expanded graph
$\HHtri$. We show that the chosen orders prevent three
edges corresponding to one queue of $\HH$ from forming
a $3$-rainbow in $\HHtri$. Thus each of the five queues
requires at most two queues, giving
$\qn(\HHtri)\le2\cdot5=10$.

We next apply the construction underlying the path bound of
\cref{lem:product-queues} to $\HHtri\boxtimes P$, using the
constructed vertex order of $\HHtri$, where $P$ records
the layers of $G$.
The inter-bag edges require at most $3\qn(\HHtri)$ queues:
the additional queue in the product construction contains
only edges between copies of the same vertex.
Intra-layer intra-bag edges use one queue, and inter-layer
intra-bag edges use two~\cite{BGR23}, giving
$\qn(G)\le3\qn(\HHtri)+1+2\le3\cdot10+3=33$.

\section{Proof of the Main Theorem}
\label{sec:bottom-parent-33}

\subsection{Preprocessing}
\label{sec:preprocessing}

To simplify tripod reordering, we first modify the recursion
of \pps so that every noninitial tripod has three nonempty
vertical paths. The modification adds vertices and edges
while preserving the BFS depths of existing vertices.

Assume that the input is a planar graph $G_0$.
Augment $G_0$ to a plane triangulation and fix a BFS tree
rooted at a vertex of the outer face, as in
\cref{sec:tripod-decomposition}.
During the tripod recursion, if a selected face produces a
tripod with one or two paths, insert a triangle
of three tree leaves inside that face and triangulate, so that
the resulting tripod has three nonempty vertical paths.
More precisely, let the selected face be \(\langle v_1,v_2,v_3\rangle\),
and let \(d(v)\) denote the BFS depth of a vertex \(v\).
For each \(j\in\{1,2,3\}\), add a leaf \(x_j\) with tree
parent \(v_j\), and add the triangle \(\langle x_1,x_2,x_3\rangle\).
For \((j,k)\in\{(1,2),(2,3),(3,1)\}\), triangulate the
quadrilateral \(\langle v_j,v_k,x_k,x_j\rangle\) with \((v_j,x_k)\) if
\(d(v_j)\geq d(v_k)\), and with \((v_k,x_j)\) otherwise.
Set \(d(x_j)=d(v_j)+1\). Every new edge joins vertices whose
depths differ by at most one, so the extended tree remains
a BFS tree and the original layers are unchanged.
Extend each nonempty vertical path to its corresponding
$x_j$; if the path is empty, use the singleton path $(x_j)$.

Henceforth, $G$ denotes the augmented graph, in which every
noninitial tripod contains three nonempty vertical paths.
It suffices to construct a queue layout of $G$, since
deleting the added vertices and edges gives a layout of
$G_0$ with the same queue bound.

\subsection{P3T for Multigraphs}
\label{sec:p3t-multigraphs}

\subsubsection{Embedded Quotient}
\label{sec:new-quotient}

In this section, we extend the \ptt algorithm described in
\cref{sec:p3t} to multigraphs, denoting the extension by
$\widehat{\ptt}$.
Contract each tripod of $G$ to a single vertex, preserving
the embedding and retaining all edges except those belonging
to its vertical paths. The resulting plane multigraph,
denoted by $\HH$, is the \df{quotient} of the tripod partition.
This graph may contain loops and parallel edges; see
\cref{fig:gap-review}(c).
We use a vertex $u\in V(\HH)$ to refer to its corresponding tripod.
An \df{incidence} at $u$ is an occurrence of an edge at $u$;
a nonloop edge contributes one incidence at each endpoint,
and a loop contributes two at its vertex.
The \df{rotation} at $u$ is the cyclic order of these
incidences in the embedding, treating parallel edges separately.
The \df{rotation} at $u$ is the
cyclic order of edge incidences in this embedding,
including both incidences of each loop and every
parallel copy.

The vertices of $\HH$ are partitioned into \df{levels}
$L_0,L_1,\ldots$ by their distance from the three initial
vertices on the outer face. These levels are distinct
from the layers $V_i$ of $G$.
We use the level components, parent and child components,
and level and binding edges defined in \cref{sec:p3t}.
Each level component is outerplane, since every vertex
has a path to the initial boundary through earlier levels.
As in the planar $3$-tree construction~\cite{ABGKP20},
every bounded face of a level component has at most three
distinct boundary vertices. Each such face contains at most
one child component; if present, its neighbors in the
preceding level are exactly the boundary vertices of that
face. Note that loops and parallel edges allow faces with one
or two boundary vertices.

Each vertical path of a tripod in $G$ corresponds to a
\df{port} at its contracted vertex in $\HH$.
Ordering the ports determines the order of the tripod's
vertices within every layer.
The following elementary lemma describes how edges are arranged around
each contracted vertex and which ports they join.

\begin{lemma}
\label{lem:new-structure}
For every vertex $u\in V(\HH)$:
\begin{enumerate}[label=(\roman*)]
    \item For distinct neighbors $v,w$ of $u$ in its
    level, the incidences of edges $(u, v)$ and $(u, w)$
    do not interleave in the rotation at $u$.

    \item The incidences belonging to each port of $u$
    are consecutive in the rotation at $u$.

    \item If a neighboring tripod $v$ is constructed after $u$
    in the \pps recursion, then $v$ has edges to at most two
    ports of $u$.
\end{enumerate}
\end{lemma}

\begin{proof}
We prove the three properties in order.
\begin{itemize}[leftmargin=*]
    \item \emph{Property (i).}
    Let $u,v,w\in L_i$. Two parallel edges $(u,v)$ bound
    a region whose interior contains no other vertex of $L_i$:
    a path from such a vertex to the initial boundary through
    decreasing levels could not cross the two edges.
    Hence no edge $(u,w)$ enters this region. All incidences
    of edges $(u,w)$ therefore lie
    in the same sector between consecutive copies of
    $(u, v)$, proving that the two groups do not interleave.

    \item \emph{Property (ii).}
    The claim is immediate for an initial tripod,
    since all its incidences belong to its only port.
    For any other tripod, a traversal around it follows both
    sides of each vertical path before proceeding to the next.
    The contraction preserves this cyclic order, so each port's
    incidences are consecutive in the rotation at $u$.

    \item \emph{Property (iii).}
    Let $p^1,p^2,p^3$
    be the vertical paths of $u$. As described in
    \cref{sec:tripod-decomposition}, constructing $u$ creates
    three recursive regions $R^1,R^2,R^3$; see
    \cref{fig:decomposition-review}(b).
    Tripods constructed inside $R^j$ have no edges to $p^j$.
    Every neighboring tripod $v$ constructed after $u$ lies in
    one of these regions, so it has edges to at most two ports
    of $u$.
\end{itemize}
\end{proof}

\subsubsection{The $\widehat{\ptt}$ Algorithm}
\label{sec:new-assignment}

As in the basic \ptt algorithm, $\widehat{\ptt}$ orders
vertices level by level, keeping each level component
consecutive and ordering its children by their containing
faces. Unlike $H$, the multigraph $\HH$ may have faces
with one or two boundary vertices as well as three.
The only difference from \ptt is therefore in the two-level
construction, which we describe through three aspects:
face assignment, face ordering, and queue assignment.

\paragraph{Face assignment.}

\begin{figure}[!tb]
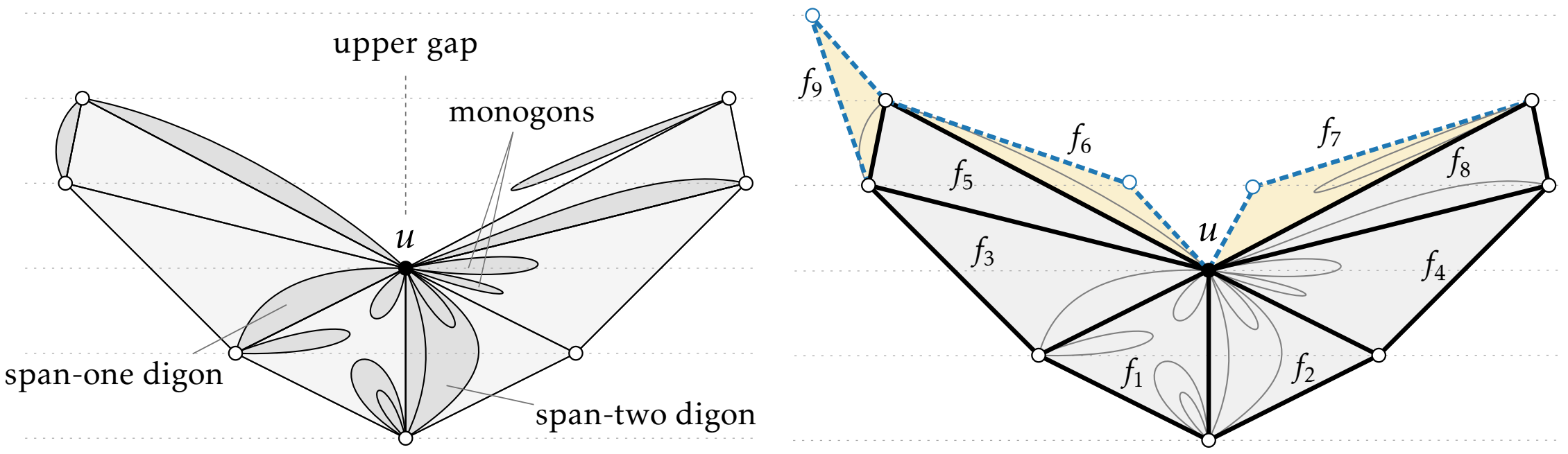

    \centering
    \begin{subfigure}[c]{.49\linewidth}
        \centering
        \includegraphics[page=2,width=\linewidth]{binding_assignment}
        \caption{Faces of $\KK$}
    \end{subfigure}
    \hfill
    \begin{subfigure}[c]{.49\linewidth}
        \centering
        \includegraphics[page=3,width=\linewidth]{binding_assignment}
        \caption{Graph $K$}
    \end{subfigure}
    \caption{Face assignment. (a) A level component $\KK$.
        (b) The graph $K$, whose faces $f_1,\ldots,f_9$ are the owners of the bounded faces of $\KK$.
        Thick edges belong to $K$; dashed blue edges and auxiliary
        vertices complete the yellow owner triangles.}
    \label{fig:face-assignment}
\end{figure}

Consider a level $L_i$ of $\HH$, and let $\KK$ be a connected
component of $\HH[L_i]$,
embedded by restricting the drawing of $\HH$ to its vertices
and edges; observe that $\KK$ is an embedded multigraph;
see \cref{fig:face-assignment}(a).
A bounded face of $\KK$ with one, two, or three distinct
boundary vertices is called a \df{monogon}, \df{digon},
or \df{triangular face}, respectively.
From the drawing of $\KK$, we construct a simple outerplane
graph $K$ by omitting loops, retaining one specified copy of
each set of parallel edges, and adding auxiliary triangles.
The construction is designed so that every bounded face of
$\KK$ lies in a triangular face of $K$, called its \df{owner}.
Next we describe the construction; see \cref{fig:face-assignment}(b) for
an illustration.

\begin{enumerate}[leftmargin=*,itemsep=2pt,parsep=0pt]
    \item \emph{Underlying simple graph.}
    We obtain $K$ from $\KK$ by deleting loops
    and retaining one copy of each set of parallel edges.

    \item \emph{Augmentation and drawing.}
    If $K$ has only one or two vertices, we add auxiliary
    vertices to form a triangle containing all bounded
    faces of $\KK$.
    Otherwise, we augment $K$ to a maximal outerplane graph,
    preserving its embedding. We then apply
    \cref{thm:outerplane-drawing} to $K$ to obtain
    an outerplane drawing in which every edge has span one or two.
    These heights also determine the spans of all nonloop
    edges of $\KK$.

    \item \emph{Parallel edges.}
    We retain the augmentation edges. Among parallel copies,
    we retain the bottommost span-one edge and the leftmost span-two
    edge, placing their digons above and to the right, respectively.
    The retained copies are shown as thick black edges in
    \cref{fig:face-assignment}(b).

    \item \emph{Digons in the outer face.}
    If a digon lies in the outer face of $K$, we add an auxiliary triangle
    above its retained span-one edge or to the right of its
    retained span-two edge. We place the new vertex one track
    above the higher endpoint in the first case, and on the
    intermediate track in the second case, using the drawing
    construction of \cref{thm:outerplane-drawing}.
    The auxiliary triangles $f_6$ and $f_9$ in
    \cref{fig:face-assignment}(b) illustrate the span-two
    and span-one cases, respectively.

    \item \emph{Monogons in the outer face.}
    Let $u$ be the vertex of a monogon in the outer face
    of $K$. We add an auxiliary triangle
    along an edge bounding the outer-face sector at $u$
    that contains the upward vertical direction.
    We choose the side containing the monogon and draw the
    triangle around it, using the same drawing construction.
    If an auxiliary triangle already exists on that side,
    we draw it to contain the monogon as well.
    The auxiliary triangle $f_7$ in
    \cref{fig:face-assignment}(b) contains such a monogon.

    \item \emph{Owner assignment.}
    We assign each bounded face of $\KK$ to the triangular
    face of $K$ containing it.
\end{enumerate}

\noindent
Thus, by construction, every bounded face of $\KK$ is contained
in a unique triangular face of $K$, and its boundary
vertices belong to that triangle.

\paragraph{Face ordering.}
A bounded face of $\KK$ is called \df{nonempty} if it contains
a child component. Our goal is to construct a linear order of all nonempty
bounded faces of $\KK$.
Recall that the original \ptt algorithm described in \cref{sec:p3t}
orders faces lexicographically by their bottom, middle, and top vertices.
We follow the same scheme: for each face of $\KK$, we identify its owner
in $K$, and then apply the \ptt algorithm to $K$.
Since several faces may have the same owner, we need a tie-breaking comparison rule for monogons, digons, and triangular faces of $\KK$.
The rule is designed so that the resulting face order agrees
with the rotation at each vertex of the owner, as explained below.

\begin{figure}[!tb]
    \centering
    \begin{subfigure}[c]{.60\linewidth}
        \centering
        \includegraphics[page=1,width=\linewidth]{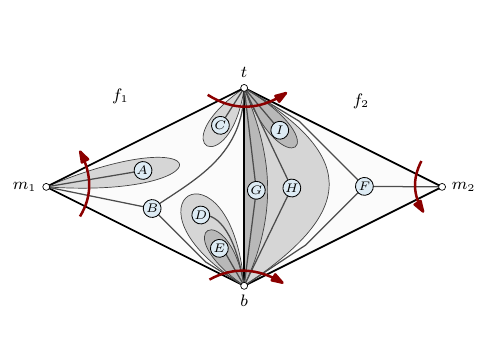}
        \caption{Faces within two owners}
        \label{fig:face-order-embedding}
    \end{subfigure}
    \hfill
    \begin{subfigure}[c]{.37\linewidth}
        \centering
        \includegraphics[page=2,width=\linewidth]{face_order}
        \caption{Local lists and common orders}
        \label{fig:face-order-lists}
    \end{subfigure}
    \caption{Ordering faces within $f_1,f_2$ from \cref{fig:face-assignment}(b).
        (a) A--I label faces; circles represent child components and gray lines binding edges.
        Arrows show the local traversals. (b) Each common order preserves all three lists;
        faces owned by $f_1$ precede those owned by $f_2$.}
    \label{fig:face-order-process}
\end{figure}

Let $\prec_K$ be the vertex order of $K$ given by
\cref{thm:outerplane-drawing}: track by track from bottom to
top, and from left to right within each track.
Faces of $\KK$ with distinct owners are ordered lexicographically
by their owners' $(b,m,t)$ triples, using $\prec_K$ for
each comparison, where $b,m,t$ are the bottom, middle,
and top vertices of the owner.
For nonempty faces of $\KK$ sharing an owner $\Delta$, we construct
three lists $\mathcal B_\Delta,\mathcal M_\Delta,\mathcal T_\Delta$;
see \cref{fig:face-order-embedding}.
At each vertex of $\Delta$, we traverse its interior angle
from one bounding edge to the other: clockwise at $b$,
and counterclockwise at $m$ and $t$.
For a specified traversal at a vertex and a specified set
of incident edges, the \df{first-edge sequence} consists
of the first edge to each distinct child component for
binding edges, or to each distinct neighbor for nonloop level edges,
listed in traversal order.
We use the binding edges to child components in faces
owned by $\Delta$. Replacing each edge of the first-edge
sequence by the face
containing its child component gives the list
$\mathcal B_\Delta$, $\mathcal M_\Delta$, or $\mathcal T_\Delta$
at $b$, $m$, or $t$, respectively.
Since each nonempty face contains one child component,
it appears at most once in each list.
We seek a linear order of the nonempty faces owned by
$\Delta$ that preserves their relative order in all three lists.

For two nonempty faces $f,g$ of $\KK$ sharing the same owner
$\Delta$, write $f\prec_{\mathcal L}g$ if both appear in the list
$\mathcal L\in\{\mathcal B_\Delta,\mathcal M_\Delta,\mathcal T_\Delta\}$
and $f$ precedes $g$ in that list.
We combine the three lists into a linear order $\prec_\Delta$
(see \cref{fig:face-order-lists}) such that, for every
$\mathcal L\in\{\mathcal B_\Delta,\mathcal M_\Delta,\mathcal T_\Delta\}$
and all $f,g\in\mathcal L$, it holds that
\(
f\prec_{\mathcal L}g \;\Longrightarrow\; f\prec_\Delta g.
\)
At each step, we choose a face that is first in every remaining
list containing it, append it to the order, and remove it
from all lists. Ties between eligible faces may be resolved
arbitrarily.

We next show that this procedure orders all nonempty faces
owned by $\Delta$. It suffices to prove that the comparisons
prescribed by the three lists contain no directed cycle:
then, whenever faces remain, at least one is first in every
remaining list containing it.
First, we show that the lists agree on their common faces.
Indeed, the clockwise
traversal at $b$ and the counterclockwise traversal at $m$
or $t$ read the region between parallel edges from the same
side, encountering its digons in the same order.
The original triangular face, if present, lies beyond the
same extreme parallel edge in both traversals.
Second, we exclude a directed cycle of comparisons.
Every digon owned by $\Delta$ has boundary vertices $b,m$
or $b,t$: a digon on edge $(m, t)$ belongs to the triangle above
that edge. A nonempty triangular face appears in all three
lists, since its child component has neighbors at all three
boundary vertices. Thus every face outside $\mathcal B_\Delta$
is a monogon at $m$ or $t$ and belongs to exactly one other
list. Any directed cycle must visit $\mathcal B_\Delta$, since
each face outside this list belongs to only one other list.
Between consecutive visits, the cycle follows comparisons
in a single list, whose order agrees with
$\mathcal B_\Delta$ on their common faces.
Thus every such segment moves strictly forward in
$\mathcal B_\Delta$, which excludes a cycle. Hence an eligible
face always exists whenever faces remain, and the procedure
defines the required order.

\paragraph{Queue assignment.}

As in the original \ptt algorithm, level edges of span one
and two belong to $Q_1$ and $Q_2$, respectively; all parallel
copies receive the same assignment.
For a child component contained in a face with owner
$\Delta$, let $b,m,t$ be the bottom, middle, and top
vertices of $\Delta$. Binding edges from $b$, $m$, and $t$
to that component belong to $\Qt$, $\Qm$, and $\Qb$,
respectively; see \cref{fig:role-intervals-queues}.
Thus a child component in a monogon or digon uses one or
two binding queues, respectively.
Loops of $\HH$ may be assigned to any queue, since they
cannot nest with another edge; their corresponding edges
in $G$ are counted separately as intra-bag edges.

\medskip
\noindent
The $\widehat{\ptt}$ algorithm applies the three steps described above
(face assignment, face ordering, and queue assignment) to
every level component of $\HH$.
As in \cref{sec:p3t}, the face orders determine the
breadth-first order of the level components; ordering the
vertices of each component $\KK$ by the restriction of
$\prec_K$ to $V(\KK)$ gives $\prec_{\HH}$.
Together with the queue assignment, this
specifies the output of $\widehat{\ptt}$.
Next, we discuss the properties of $\widehat{\ptt}$.

\subsubsection{Properties of the $\widehat{\ptt}$ Algorithm}
\label{sec:p3t-properties}

\begin{figure}[tb]
    \centering
    \begin{subfigure}[t]{.49\linewidth}
        \centering
        \includegraphics[page=1,width=\linewidth]{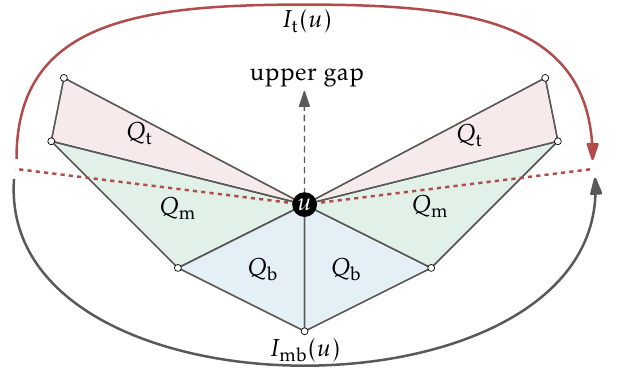}
        \caption{Rotation intervals at $u$}
        \label{fig:role-intervals-embedding}
    \end{subfigure}
    \hfill
    \begin{subfigure}[t]{.49\linewidth}
        \centering
        \includegraphics[page=2,width=\linewidth]{role_intervals}
        \caption{Binding edges within two owners}
        \label{fig:role-intervals-queues}
    \end{subfigure}
    \caption{Queue assignment: red, green, and blue indicate $\Qt,\Qm,\Qb$.
        (a) $I_{\mathrm t}(u)$ includes its boundary level edges and is read clockwise;
        $I_{\mathrm{mb}}(u)$ is read counterclockwise.
        (b) Binding edges within the owners from \cref{fig:face-order-embedding};
        interior circles represent child components.}
    \label{fig:role-intervals}
\end{figure}

Here we establish the five-queue bound of $\widehat{\ptt}$ and show
that the resulting vertex order $\prec_{\HH}$ is compatible
with the rotation system around each vertex $u$ of $\HH$.
To express this relation, we define two intervals in the
rotation at each vertex, using the augmented graph~$K$;
see \cref{fig:role-intervals-embedding}.
At each vertex $u\in V(\HH)$, we consider the rotation at
$u$ in $\HH$, restricted to level edges and binding edges
to child components. Each loop appears twice, once for
each incidence. An \df{interval} is an ordered list of
consecutive incidences in this rotation.
Starting from the horizontal direction pointing left,
we read the rotation of $K$ at $u$ clockwise.
Let $e_{\mathrm L}$ and $e_{\mathrm R}$ be the first and
last edges encountered whose other endpoints lie above $u$.
We define:
\begin{itemize}[leftmargin=*]
    \item \df{$I_{\mathrm t}(u)$} consists of the incidences
    of $\HH$ encountered clockwise from $e_{\mathrm L}$
    to $e_{\mathrm R}$, including all copies of
    $e_{\mathrm L}$ and $e_{\mathrm R}$ and the incidences
    between those copies.

    \item \df{$I_{\mathrm{mb}}(u)$} consists of the remaining
    incidences of $\HH$, read counterclockwise from $e_{\mathrm L}$
    to $e_{\mathrm R}$.
\end{itemize}
If there is only one edge to a vertex above $u$, the two
boundary edges coincide and $I_{\mathrm t}(u)$ contains
all copies of that edge and the incidences between them.
If there is no such edge, $I_{\mathrm t}(u)$ is empty
and $I_{\mathrm{mb}}(u)$ is the entire cyclic list, read
counterclockwise starting from the upward vertical direction.
Auxiliary edges determine the boundaries but do not appear
in either interval.
All first-edge sequences follow these traversals.
A first-edge sequence is \df{compatible} with $\prec_{\HH}$
if its opposite endpoints occur in $\prec_{\HH}$-order.
The lemma establishes this compatibility for the first-edge
sequences in the two intervals.

\begin{lemma}
    \label{lem:multi-p3t}
    Let $\HH$ be the quotient of a preprocessed plane
    triangulation. The $\widehat{\ptt}$ algorithm gives a $5$-queue
    layout of $\HH$ with vertex order $\prec_{\HH}$ and queues
    $Q_1,Q_2,\Qt,\Qm,\Qb$. Moreover:
    \begin{enumerate}[label=(\roman*)]
        \item For each binding queue and each child component,
        all edges of that queue joining the component to the
        preceding level have the same preceding-level endpoint.

        \item For each $u\in V(\HH)$, the incidences at $u$ of
        $\Qt$-edges to the next level lie in $I_{\mathrm t}(u)$,
        and those of $\Qm$- and $\Qb$-edges lie in
        $I_{\mathrm{mb}}(u)$.
        For each of these queues, the first-edge sequence
        in the corresponding interval is compatible with
        $\prec_{\HH}$.

        \item For each $u\in V(\HH)$ and each $Q\in\{Q_1,Q_2\}$,
        the first-edge sequence of $Q$ in each interval is compatible
        with $\prec_{\HH}$.
    \end{enumerate}
\end{lemma}

\begin{proof}

We first establish the five-queue bound and property~(i),
and then prove the rotation properties (ii) and (iii).

\begin{itemize}[leftmargin=*]
    \item \emph{Queue bound and property~(i).}
    The span-one and span-two level edges form queues $Q_1,Q_2$
    by \cref{thm:outerplane-drawing}; parallel copies cannot nest
    because they share endpoints. Concatenating the orders of
    level components and levels preserves these two queues.

    For binding edges, the bottom, middle, and top vertices of
    the owners are each nondecreasing in $\prec_K$ along their
    lexicographic $(b,m,t)$-order, as in the original
    \ptt algorithm described in \cref{sec:p3t}.
    Each binding queue uses at most one vertex of an owner
    as its preceding-level endpoint, proving~(i).
    Faces with the same owner therefore use the same
    preceding-level endpoint in each binding queue. Thus, in each of
    $\Qt,\Qm,\Qb$, the preceding-level endpoints are nondecreasing
    along the child order (the order of child components induced
    by $\prec_{\HH}$). Since each child component occupies
    a consecutive block, edges from distinct preceding-level
    endpoints cannot nest; edges from the same endpoint cannot
    nest either. Children of different parent components follow
    their parents' order, and edges joining different consecutive
    level pairs cannot nest. Hence all three binding sets are
    queues.

    \item \emph{Property (ii).}
    Fix $u\in V(\HH)$ and $Q\in\{\Qt,\Qm,\Qb\}$. We first verify
    interval containment.

    Let $f$ be a nonempty bounded face of $\KK$ whose child
    component is joined to $u$, and let $\Delta$ be the owner
    of $f$ in $K$.
    Since $f$ lies inside $\Delta$, the binding edges from $u$
    to that child component occur within the interior angle
    of $\Delta$ at $u$.
    If $u$ is the bottom vertex of $\Delta$, this angle lies
    between two edges leading to higher vertices and hence
    within $I_{\mathrm t}(u)$. If $u$ is the middle or top
    vertex of $\Delta$, the angle lies outside that interval
    and hence within $I_{\mathrm{mb}}(u)$.
    By the queue assignment, the incidences of $\Qt$-edges
    to the next level therefore lie in $I_{\mathrm t}(u)$,
    and those of $\Qm$- and $\Qb$-edges lie in
    $I_{\mathrm{mb}}(u)$.

    It remains to prove that the first-edge sequence of $Q$
    in its corresponding interval is compatible with
    $\prec_{\HH}$.
    For faces sharing an owner $\Delta$, their order
    $\prec_\Delta$ agrees with the order of their first
    binding edges in the prescribed traversal at $u$,
    by construction.
    Since the binding edges belong to one queue $Q$, $u$ has
    the same role (bottom, middle, or top) in all their owners.
    If $u$ is their bottom vertex, clockwise traversal encounters
    the edges to their middle vertices in left-to-right order,
    since those vertices lie on one track.
    If two owners share the middle vertex, they share an edge
    with $u$ and lie on opposite sides of it; the traversal
    therefore encounters them in the left-to-right order of
    their top vertices.
    If $u$ is their middle or top vertex, counterclockwise
    traversal similarly encounters the edges to their bottom
    vertices in left-to-right order. When the bottom vertex
    is shared, the owners lie on opposite sides of their common
    edge, and the traversal encounters them in the left-to-right
    order of their remaining vertices.
    Thus, in each case, the traversal follows the owners'
    lexicographic $(b,m,t)$-order.
    Since child components occur consecutively in $\prec_{\HH}$
    in the order of their containing faces, the first-edge
    sequence of $Q$ is compatible with $\prec_{\HH}$.

    \item \emph{Property (iii).}
    Fix $u\in V(\HH)$ and $Q\in\{Q_1,Q_2\}$.
    All nonloop edges of $Q$ have the same span. Thus the opposite
    endpoints above $u$ have a common height, and their clockwise
    order along $I_{\mathrm t}(u)$ is their left-to-right
    order, which agrees with $\prec_{\HH}$.
    Similarly, the opposite endpoints below $u$ lie on one
    track, and counterclockwise traversal of
    $I_{\mathrm{mb}}(u)$ encounters them from left to right,
    again agreeing with $\prec_{\HH}$.
    Parallel copies do not interleave with edges to
    other neighbors by \cref{lem:new-structure}(i),
    so retaining only the first edge to each neighbor
    preserves these orders.
\end{itemize}

\end{proof}

\subsection{Tripod Reordering}
\label{sec:new-path-reordering}

By \cref{lem:multi-p3t}, $\HH$ has a $5$-queue layout
with vertex order $\prec_{\HH}$.
Following the approach outlined in \cref{sec:overview},
we now expand each noninitial tripod vertex of $\HH$
into one vertex for each of its three ports.
The resulting graph is the \df{expanded quotient},
denoted $\HHtri$.
Each edge of $\HH$ joining distinct tripods becomes an
edge between the corresponding ports, and parallel edges
are retained.
We keep the ports of each tripod consecutive, with
the tripods ordered by $\prec_{\HH}$.
With arbitrary port orders, the allocation underlying
\cref{lem:product-queues} gives
$\qn(\HHtri)\le 3\qn(\HH)\le 3\cdot5=15$.
Using \cref{lem:multi-p3t},
we choose port orders yielding $\qn(\HHtri)\le10$,
as proved in \cref{sec:new-compatibility}.

\begin{figure}[tb]
    \centering
    \begin{subfigure}[c]{.49\linewidth}
        \centering
        \includegraphics[page=1,width=\linewidth]{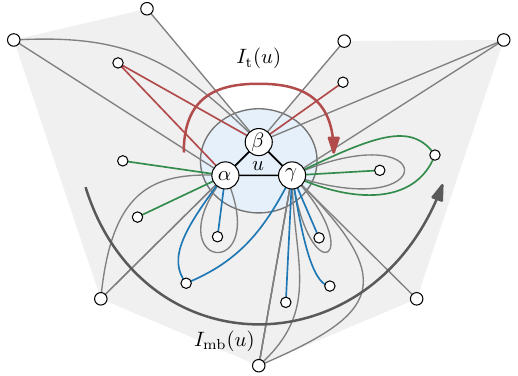}
        \caption{The ports and rotation at $u$}
        \label{fig:port-reordering-rotation}
    \end{subfigure}
    \hfill
    \begin{subfigure}[c]{.49\linewidth}
        \centering
        \includegraphics[page=2,width=\linewidth]{port_reordering}
        \caption{The resulting port order}
        \label{fig:port-reordering-order}
    \end{subfigure}
    \caption{Tripod reordering when $I_{\mathrm t}(u)$ meets all three
    ports. Clockwise traversal of $I_{\mathrm t}(u)$ first encounters
    the ports in the order $\alpha,\beta,\gamma$; their chosen order
    is $\beta\prec_u\alpha\prec_u\gamma$.
    Red, green, and blue binding edges belong to $\Qt$, $\Qm$,
    and $\Qb$, respectively.}
    \label{fig:port-reordering}
\end{figure}

For each $u\in V(\HH)$, an interval \df{meets} a port
of $u$ if it contains an incidence belonging to that port;
see \cref{fig:port-reordering-rotation}.
The incidences of each port are consecutive in the rotation
at $u$ by \cref{lem:new-structure}(ii), and remain consecutive
after restricting to level edges and binding edges to child
components. If both $I_{\mathrm t}(u)$ and
$I_{\mathrm{mb}}(u)$ met all three ports, their two boundary
positions would split all three port blocks, which is
impossible. Thus at most one interval meets all three ports.
For a tripod with three ports, we use the names
$\alpha,\beta,\gamma$, assigned by the traversals below.
We choose the order $\prec_u$ of the ports of $u$ as follows:
\begin{itemize}
    \item If $I_{\mathrm t}(u)$ meets the three ports,
    we read it clockwise from its first incidence.
    We name the first encountered port $\alpha$, the next
    distinct port $\beta$, and the remaining port $\gamma$,
    and set $\beta\prec_u\alpha\prec_u\gamma$.

    \item Otherwise, if $I_{\mathrm{mb}}(u)$ meets the three ports,
    we read it counterclockwise from its first incidence.
    We name the first port $\alpha$, the next
    distinct port $\beta$, and the remaining port $\gamma$,
    and set $\beta\prec_u\alpha\prec_u\gamma$.

    \item If neither interval meets the three ports,
    we choose an arbitrary order of the ports.
\end{itemize}
Ordering the vertices of $\HHtri$ first by $\prec_{\HH}$
and then by the orders $\prec_u$ within each tripod gives
the \df{port order} $\prec_{\HHtri}$ of $\HHtri$;
see \cref{fig:port-reordering-order}.
The following lemma relates the local port orders to the
orders of neighboring tripods and child components.

\begin{lemma}
\label{lem:new-admissible-ports}
Let $u\in V(\HH)$ have three ports ordered
$\beta\prec_u\alpha\prec_u\gamma$.
Let $(u,v_{\alpha})$, $(u,v_{\beta})$, and $(u,v_{\gamma})$
be edges of one queue $Q$ in the layout given by
\cref{lem:multi-p3t}, belonging at $u$ to ports
$\alpha,\beta,\gamma$, respectively.
\begin{enumerate}[label=(\roman*)]
    \item If $Q\in\{Q_1,Q_2\}$ and $v_{\alpha},v_{\beta},v_{\gamma}$ lie
    on the same side of $u$ in $\prec_{\HH}$, then
    $v_{\beta}\ne v_{\gamma}$ implies $v_{\beta}\prec_{\HH}v_{\gamma}$.

    \item If $Q\in\{\Qt,\Qm,\Qb\}$, $u\in L_i$, and
    $v_{\alpha},v_{\beta},v_{\gamma}\in L_{i+1}$, let $C_{\alpha},C_{\beta},C_{\gamma}$ be their
    level components. Then $C_{\gamma}\preceq C_{\alpha}\preceq C_{\beta}$
    is impossible in the child order.
\end{enumerate}
\end{lemma}

\begin{proof}
Whenever the three incidences at $u$ lie in one interval,
that interval meets all three ports and determines
$\prec_u$. Its traversal encounters the ports in the
order $\alpha,\beta,\gamma$, possibly followed by
$\alpha$ again. In particular, every incidence belonging
to $\beta$ precedes every incidence belonging to $\gamma$.

\begin{itemize}[leftmargin=*]
\item \emph{Property (i).} Let $\KK$ be the level component
containing $u$, and let $K$ be its augmented graph.
Since the three edges belong to one queue
$Q\in\{Q_1,Q_2\}$, they have the same span.
Since $v_{\alpha},v_{\beta},v_{\gamma}$ lie on the same
side of $u$ in $\prec_{\HH}$, they lie on one track,
either above or below $u$.
Their incidences at $u$ therefore all belong to
$I_{\mathrm t}(u)$ or all to $I_{\mathrm{mb}}(u)$,
respectively.
If $v_{\beta}\ne v_{\gamma}$, the edges from $u$ to these two neighbors
occur in noninterleaving groups by \cref{lem:new-structure}(i),
and the interval boundaries do not split the edges to either neighbor. Thus the
group leading to $v_{\beta}$ precedes the group leading to $v_{\gamma}$.
Their first occurrences follow $\prec_{\HH}$ by
\cref{lem:multi-p3t}(iii), so $v_{\beta}\prec_{\HH}v_{\gamma}$.

\item \emph{Property (ii).} The three incidences lie in
$I_{\mathrm t}(u)$ when $Q=\Qt$, and in
$I_{\mathrm{mb}}(u)$ when $Q\in\{\Qm,\Qb\}$, by
\cref{lem:multi-p3t}(ii).
We first show that the binding edges of one child
component cannot span all incidences of an intervening
port in the rotation at $u$.
Consider a child component $C$ joined to $u$ by at least two edges of $Q$.
Two distinct binding edges together with a simple path in $C$
form a cycle, possibly a digon, whose vertices other than
$u$ lie in $L_{i+1}$. Its bounded side contains no other
vertex of level at most $i$: a path from such a vertex to
the initial boundary through decreasing levels could not
cross the cycle. Hence its bounded sector at $u$ contains
no nonloop incidence to level at most $i$. The cycle lies inside the owner of the face containing
$C$. Its bounded sector at $u$ therefore lies within
the owner's interior angle. By the interval definitions,
this angle lies entirely within one interval, so neither
interval boundary lies inside it. If $u$ has no neighbor
above it in $K$, the traversal of $I_{\mathrm{mb}}(u)$ starts
upward, outside the owner's interior angle. Thus the traversal starts outside the
bounded sector. Applied to the first and last binding
edges to $C$, this shows that the sector traversed
between them is the bounded one. Since $u$ has three ports, it represents a noninitial tripod.
By the \pps construction described in
\cref{sec:tripod-decomposition}, each vertical path has a tree
edge to a boundary tripod of the region in which $u$ was created.
These boundary tripods are pairwise adjacent and separate
$u$ from the initial vertices. If their minimum level is
$d$, then $u$ has level $i=d+1$, and every boundary tripod
has level at most $d+1=i$.
Thus each port has an edge to a vertex of level at most $i$.
Here we consider the full rotation at $u$, including
edges to the preceding level. The incidence of such
an edge lies outside the bounded sector. Since each
port occupies a consecutive block in the full rotation,
the sector cannot contain an entire port block.

In the following argument, a child component meets a
port if an edge of $Q$ joins that component to the port.
This restriction partitions the child components joined
to $u$ by edges of $Q$ into three consecutive groups
in the child order.
Children whose
first binding edge has its incidence among the initial $\alpha$-incidences meet
only $\alpha,\beta$: reaching $\gamma$ would span all $\beta$-incidences.
Similarly, children whose first binding edge has its incidence at $\beta$
meet only $\beta,\gamma$, and those whose first binding edge has its incidence
at $\gamma$ or among the terminal $\alpha$-incidences meet only $\gamma,\alpha$.
These three groups occur in this order in the child order
by \cref{lem:multi-p3t}(ii).

Suppose, for a contradiction, that
$C_{\gamma}\preceq C_{\alpha}\preceq C_{\beta}$
in the child order.
Since $C_{\gamma}$ meets $\gamma$, it belongs to the second or third
group, so $C_{\alpha}$ cannot belong to the first group. Since
$C_{\alpha}$ meets $\alpha$, it cannot belong to the second group
either. Thus $C_{\alpha}$ belongs to the third group. But
$C_{\alpha}\preceq C_{\beta}$, and no child in that group meets $\beta$,
a contradiction.
\end{itemize}
\end{proof}

\subsection{Counting Queues}
\label{sec:new-compatibility}

Recall that $\HHtri$ is obtained by expanding each
noninitial tripod vertex of $\HH$ into its ordered
triple of ports.
We first show that the resulting port order gives a
$10$-queue layout of $\HHtri$, and then derive the bound
for $G$.

\begin{lemma}
\label{lem:new-path-order}
Let $\HH$ be the quotient of a preprocessed plane
triangulation, and let $\HHtri$ be its expanded quotient.
Then $\HHtri$ contains no $11$-rainbow in the port order $\prec_{\HHtri}$.
\end{lemma}

\begin{proof}
For each queue $Q\in\{Q_1,Q_2,\Qt,\Qm,\Qb\}$ of $\HH$,
we show that the corresponding edges of $\HHtri$ contain
no $3$-rainbow in $\prec_{\HHtri}$.
Since these five queues partition the edges of $\HH$,
every rainbow in $\HHtri$ then has at most $2\cdot5=10$
edges, proving the claim.

Suppose, for a contradiction, that
three such edges form a $3$-rainbow.
Let $e_1,e_2,e_3$ be the edges of the $3$-rainbow, and let
$f_1,f_2,f_3$ be the corresponding edges in $\HH$, all belonging to $Q$.
For $i\ne j$, the edges $f_i$ and $f_j$ share an endpoint;
otherwise they would nest in $\prec_{\HH}$. If the three edges have
no common endpoint, they form a triangle on vertices
$u\prec_{\HH}v\prec_{\HH}w$. Of the two edges among
$e_1,e_2,e_3$ corresponding to $(u, v)$ and $(v, w)$, the former
has the first endpoint in $\prec_{\HHtri}$ and the latter has the
last endpoint. Hence they cannot nest, a contradiction.
Thus $f_1,f_2,f_3$ share a vertex $u$, and the corresponding
endpoints are distinct ports of tripod $u$. All opposite
endpoints belong to tripods on the same side of $u$ in
$\prec_{\HH}$, since edges to opposite sides cannot nest.
Name the ports of $u$ so that
$\beta\prec_u\alpha\prec_u\gamma$, and let
$y_\alpha,y_\beta,y_\gamma$ be their opposite endpoints.
Nesting requires
$y_\gamma\prec_{\HHtri}y_\alpha\prec_{\HHtri}y_\beta$.

\begin{itemize}[leftmargin=*]
\item \emph{Level edges: $Q\in\{Q_1,Q_2\}$.}
If $y_{\beta},y_{\gamma}$
are ports of distinct tripods $v,w$, then
\cref{lem:new-admissible-ports}(i) gives $v\prec_{\HH}w$,
and hence $y_{\beta}\prec_{\HHtri} y_{\gamma}$, a contradiction. Otherwise,
all three opposite endpoints are ports of one tripod,
since its ports are consecutive in $\prec_{\HHtri}$. By
\cref{lem:new-structure}(iii), the earlier-created of the
two tripods is either initial or has a port missed
by the later tripod. Hence at most two of its ports can
be endpoints of edges between the two tripods,
contradicting the three independent edges.

\item \emph{Binding edges: $Q\in\{\Qt,\Qm,\Qb\}$.}
If $u$ lies in the later level, \cref{lem:multi-p3t}(i)
puts all three opposite endpoints in one parent tripod.
All six endpoints then belong to two tripods, contradicting
\cref{lem:new-structure}(iii) as in the level-edge case.
Thus $u$ is the preceding-level endpoint of $f_1,f_2,f_3$.
Let $C_{\alpha},C_{\beta},C_{\gamma}$ be the child
components containing the tripods of $y_{\alpha},y_{\beta},y_{\gamma}$,
respectively. The ports belonging to each child component
are consecutive, so $C_{\gamma}\preceq C_{\alpha}\preceq C_{\beta}$ in the
child order, contradicting \cref{lem:new-admissible-ports}(ii).
\end{itemize}

In both cases we obtain a contradiction, which completes the proof.
\end{proof}

\maintheorem*

\begin{proof}
Without loss of generality, we may assume that $G$ is a plane
triangulation with the preprocessing of
\cref{sec:preprocessing} applied.
To build a queue layout of $G$, we follow the \pps construction
outlined in \cref{sec:layered-partitions}, retaining loops
and parallel edges in the quotient $\HH$.
The $\widehat{\ptt}$ algorithm gives a $5$-queue layout
of $\HH$ by \cref{lem:multi-p3t}.
By \cref{lem:new-path-order}, the port order
$\prec_{\HHtri}$ gives a $10$-queue layout of $\HHtri$.
Since each port corresponds to a vertical path meeting
each layer at most once, the inter-bag subgraph of $G$
is contained in $\HHtri\boxtimes P$, where $P$ records
the layers.
Applying the construction underlying the path bound of
\cref{lem:product-queues} to this $10$-queue layout gives
a $31$-queue layout of $\HHtri\boxtimes P$, with vertices
ordered first by layer and then by $\prec_{\HHtri}$.
Each queue of $\HHtri$ contributes one queue for edges
within a layer and two queues for edges between consecutive
layers, one for each direction of the port order.
The additional queue in this construction contains all edges
between copies of the same port. Since no inter-bag edge
joins vertices of the same port, the inter-bag edges use
at most $3\cdot10=30$ queues.

In this layout, the vertices of each noninitial tripod
in each layer form a consecutive block of size three, while
initial tripods contribute one vertex each;
so intra-layer intra-bag edges form one queue.
Inter-layer intra-bag edges use two queues~\cite{BGR23}.
Indeed, a $3$-rainbow would lie in one tripod and two
consecutive layers. Since the port order is fixed,
its outermost and innermost edges would join the first
and third vertical paths in opposite directions.
Both edges lie in the same recursive region, where their
endpoints alternate along the boundary, contradicting planarity.
Hence $\qn(G)\le 30+1+2=33$.
\end{proof}

\section{Conclusions}
\label{sec:conclusions}

As a conclusion, we note that our tripod reordering
construction also yields an improved bound for bipartite
planar graphs by the argument of~\cite{FKMPR23}.
We include the short proof for completeness.

\bitheorem*

\begin{proof}
Let $G$ be a bipartite planar graph. Augment $G$ to a plane
quadrangulation and choose a layering.
Triangulating its faces with diagonals joining vertices
in the same layer and applying the preprocessing of
\cref{sec:preprocessing} preserves the layers of all
existing vertices. The proof of \cref{thm:main} gives a
queue layout of the resulting graph. Deleting the added
vertices and edges gives a queue layout of $G$.

Since $G$ is bipartite, it has no intra-layer edges.
Thus only the two inter-layer queues for each queue of
$\HHtri$ are needed in the queue layout of $\HHtri\boxtimes P$.
By \cref{lem:new-path-order}, the inter-bag edges therefore
use at most $2\cdot10=20$ queues. The inter-layer intra-bag
edges use two queues, as in the proof of \cref{thm:main}.
Hence $\qn(G)\le  2\cdot10 + 2 = 22$.
\end{proof}

Substantial gaps remain between the upper and lower bounds.
The largest queue number of a planar graph lies between
$4$~\cite{ABGKP20} and $33$, while for bipartite planar graphs
it lies between $3$~\cite{FKMPR23} and $22$.
Determining either value remains an open problem.
Even for planar $3$-trees, the largest queue number is
only known to be either four or five~\cite{ABGKP20}.

\paragraph{Declaration of generative AI use.}
The proof strategy for \cref{thm:main} and the main ideas were
developed manually by the author following the identification of the gap in \cite{PupGap26}.
The author used AI tools,
such as ChatGPT Codex (Sol/Astra) and Claude Code (Opus/Fable),
for checking proof arguments, proofreading, editing the manuscript, and
implementing the algorithm, following the author's earlier prototype
of the $49$-queue layout algorithm for planar graphs.

\paragraph{Source code.}
A full implementation of the $33$-queue layout algorithm
for planar graphs presented in \cref{sec:bottom-parent-33}
is available in~\cite{bob}.

\bibliographystyle{abbrv}
\bibliography{refs}

@book{AZ10,
    author    = {Martin Aigner and G{\"u}nter M. Ziegler},
    title     = {Proofs from {The Book}},
    edition   = {4},
    publisher = {Springer},
    year      = {2010}
}

@article{DF15,
	author    = {Vida Dujmovi{\'c} and
	Fabrizio Frati},
	title     = {Stack and Queue Layouts via Layered Separators},
	journal   = {J. Graph Algorithms Appl.},
	volume    = {22},
	number    = {1},
	pages     = {89--99},
	year      = {2018},
}

@article{DPW04,
    title={Track Layouts of Graphs},
    author={Dujmovi{\'c}, Vida and P{\'o}r, Attila and Wood, David R},
    journal={Discrete Mathematics \& Theoretical Computer Science},
    volume={6},
    number={2},
    pages={497--522},
    year={2004}
}

@article{Duj15,
    title={Graph layouts via layered separators},
    author={Dujmovi{\'c}, Vida},
    journal={Journal of Combinatorial Theory, Series B},
    volume={110},
    pages={79--89},
    year={2015},
    publisher={Elsevier}
}

@article{HR92,
    title={Laying out graphs using queues},
    author={Heath, Lenwood S and Rosenberg, Arnold L},
    journal={{SIAM} J. Comput.},
    volume={21},
    number={5},
    pages={927--958},
    year={1992},
    publisher={SIAM}
}

@article{HLR92,
    title={Comparing queues and stacks as machines for laying out graphs},
    author={Heath, Lenwood S and Leighton, Frank Thomson and Rosenberg, Arnold L},
    journal={{SIAM} J. Discrete Math.},
    volume={5},
    number={3},
    pages={398--412},
    year={1992},
    publisher={SIAM}
}

@article{Wie17,
    author    = {Veit Wiechert},
    title     = {On the Queue-Number of Graphs with Bounded Tree-Width},
    journal   = {Electr. J. Comb.},
    volume    = {24},
    number    = {1},
    pages     = {P1.65},
    year      = {2017},
}

@article{DMW05,
    title={Layout of graphs with bounded tree-width},
    author={Dujmovi{\'c}, Vida and Morin, Pat and Wood, David R},
    journal={{SIAM} J. Comput.},
    volume={34},
    number={3},
    pages={553--579},
    year={2005},
    publisher={SIAM}
}

@phdthesis{Pem92,
    title={Exploring the powers of stacks and queues via graph layouts},
    author={Pemmaraju, Sriram V},
    year={1992},
    school={Virginia Tech}
}

@article{DW05,
    title={Stacks, Queues and Tracks: Layouts of Graph Subdivisions},
    author={Dujmovi{\'c}, Vida and Wood, David R},
    journal={Discrete Mathematics and Theoretical Computer Science},
    volume={7},
    pages={155--202},
    year={2005}
}

@article{BFP13,
    title={On the queue number of planar graphs},
    author={Di Battista, Giuseppe and Frati, Fabrizio and Pach, Janos},
    journal={{SIAM} J. Comput.},
    volume={42},
    number={6},
    pages={2243--2285},
    year={2013},
    publisher={SIAM}
}

@article{ABGKP20,
    author       = {Jawaherul Md. Alam and
    Michael A. Bekos and
    Martin Gronemann and
    Michael Kaufmann and
    Sergey Pupyrev},
    title        = {Queue Layouts of Planar 3-Trees},
    journal      = {Algorithmica},
    volume       = {82},
    number       = {9},
    pages        = {2564--2585},
    year         = {2020},
    doi          = {10.1007/S00453-020-00697-4},
}

@inproceedings{Pup17,
	author    = {Sergey Pupyrev},
	title     = {Mixed Linear Layouts of Planar Graphs},
	booktitle = {International Symposium on Graph Drawing and Network Visualization},
	series    = {LNCS},
	volume    = {10692},
	pages     = {197--209},
	publisher = {Springer},
	year      = {2017}
}

@article{DJMMUW20,
    title={Planar graphs have bounded queue-number},
    author={Dujmovi{\'c}, Vida and Joret, Gwena{\"e}l and Micek, Piotr and Morin, Pat and Ueckerdt, Torsten and Wood, David R},
    journal={Journal of the ACM},
    volume={67},
    number={4},
    pages={1--38},
    year={2020},
    publisher={ACM New York, NY, USA},
    doi = {10.1145/3385731}
}

@article{BGR23,
    title={An improved upper bound on the queue number of planar graphs},
    author={Bekos, Michael A and Gronemann, Martin and Raftopoulou, Chrysanthi},
    journal={Algorithmica},
    volume={85},
    number={2},
    pages={544--562},
    year={2023},
    publisher={Springer}
}

@article{W05,
    title={Queue layouts of graph products and powers},
    author={Wood, David R},
    journal={Discrete Mathematics \& Theoretical Computer Science},
    volume={7},
    year={2005},
    publisher={Episciences.org}
}

@InProceedings{FKMPR23,
    author="F{\"o}rster, Henry
    and Kaufmann, Michael
    and Merker, Laura
    and Pupyrev, Sergey
    and Raftopoulou, Chrysanthi",
    editor="Morin, Pat and Suri, Subhash",
    title="Linear Layouts of Bipartite Planar Graphs",
    booktitle="Algorithms and Data Structures",
    year="2023",
    publisher="Springer Nature Switzerland",
    address="Cham",
    pages="444--459",
}

@inproceedings{Katheder0PU25,
    author       = {Julia Katheder and
    Michael Kaufmann and
    Sergey Pupyrev and
    Torsten Ueckerdt},
    title        = {Transforming Stacks into Queues: Mixed and Separated Layouts of Graphs},
    booktitle    = {STACS},
    series       = {LIPIcs},
    volume       = {327},
    pages        = {56:1--56:18},
    publisher    = {Schloss Dagstuhl - Leibniz-Zentrum f{\"{u}}r Informatik},
    year         = {2025},
    doi          = {10.4230/LIPICS.STACS.2025.56},
}

@article{AlamBGKP22,
    author       = {Jawaherul Md. Alam and
    Michael A. Bekos and
    Martin Gronemann and
    Michael Kaufmann and
    Sergey Pupyrev},
    title        = {The mixed page number of graphs},
    journal      = {Theor. Comput. Sci.},
    volume       = {931},
    pages        = {131--141},
    year         = {2022},
    doi          = {10.1016/J.TCS.2022.07.036},
}

@article{ABGKP23,
    author       = {Jawaherul Md. Alam and
    Michael A. Bekos and
    Martin Gronemann and
    Michael Kaufmann and
    Sergey Pupyrev},
    title        = {Lazy Queue Layouts of Posets},
    journal      = {Algorithmica},
    volume       = {85},
    pages        = {1176--1201},
    year         = {2023},
    doi          = {10.1007/s00453-022-01067-y},
}

@inproceedings{Pup22,
    author       = {Sergey Pupyrev},
    editor       = {Patrizio Angelini and Reinhard von Hanxleden},
    title        = {Queue Layouts of Two-Dimensional Posets},
    booktitle    = {International Symposium on Graph Drawing and Network Visualization},
    series       = {LNCS},
    volume       = {13764},
    pages        = {353--360},
    publisher    = {Springer},
    year         = {2022},
    doi          = {10.1007/978-3-031-22203-0\_25},
}

@article{PupGap26,
    author       = {Sergey Pupyrev},
    title        = {A Gap in the 42-Queue Layout Algorithm for Planar Graphs},
    journal      = {CoRR},
    volume       = {abs/2608.05508},
    year         = {2026},
    note         = {arXiv:2608.05508},
    eprinttype   = {arXiv},
    eprint       = {2608.05508},
}

@article{BFGMMRU19,
    author       = {Michael A. Bekos and
    Henry F{\"{o}}rster and
    Martin Gronemann and
    Tamara Mchedlidze and
    Fabrizio Montecchiani and
    Chrysanthi N. Raftopoulou and
    Torsten Ueckerdt},
    title        = {Planar Graphs of Bounded Degree Have Bounded Queue Number},
    journal      = {{SIAM} J. Comput.},
    volume       = {48},
    number       = {5},
    pages        = {1487--1502},
    year         = {2019},
    doi          = {10.1137/19M125340X},
}

@misc{bob,
    author       = {Pupyrev, Sergey},
    title        = {A {SAT}-based solver for constructing optimal linear layouts of graphs},
    howpublished = {\url{https://github.com/spupyrev/bob}},
    year         = {2017},
    note         = {Accessed: November 9, 2026}
}

\end{document}